\documentclass[prl,twocolumn]{revtex4-1}

\usepackage[T1]{fontenc}
\usepackage[latin9]{inputenc}
\usepackage{amsmath}
\usepackage{amssymb}
\usepackage{graphicx}
\usepackage{float}
\usepackage[colorlinks]{hyperref}
\usepackage{xcolor}
\usepackage{adjustbox}
\usepackage{ulem}
\makeatletter

\newcommand{\beginsupplement}{%
        \setcounter{table}{0}
        \renewcommand{\thetable}{S\arabic{table}}%
        \setcounter{figure}{0}
        \renewcommand{\thefigure}{S\arabic{figure}}%
        \setcounter{section}{0}
        \renewcommand{\thesection}{S\arabic{section}}%
        \setcounter{section}{0}
        \renewcommand{\thesection}{S\arabic{section}}%
        \setcounter{subsection}{0}
        \renewcommand{\thesubsection}{S\arabic{section}.\arabic{subsection}}%
        \setcounter{equation}{0}
        \renewcommand{\theequation}{S\arabic{equation}}%
     }
\makeatother

\global\long\def\ket#1{\left| #1\right\rangle }

\global\long\def\av#1{\left\langle #1 \right\rangle }

\definecolor{applegreen}{rgb}{0.55, 0.71, 0.0}

\begin{document}

\title{Classical and quantum liquids induced by quantum fluctuations}

\author{Miguel M. Oliveira}
\email{miguel.m.oliveira@tecnico.ulisboa.pt}
\affiliation{CeFEMA, Instituto Superior T\'ecnico, Universidade de Lisboa Av. Rovisco
Pais, 1049-001 Lisboa, Portugal}

\author{Pedro Ribeiro}
\email{ribeiro.pedro@gmail.com}
\affiliation{CeFEMA, Instituto Superior T\'ecnico, Universidade de Lisboa Av. Rovisco
Pais, 1049-001 Lisboa, Portugal}
\affiliation{Beijing Computational Science Research Center, Beijing 100193, China}

\author{Stefan Kirchner}
\email{stefan.kirchner@correlated-matter.com}
\affiliation{Zhejiang Institute of Modern Physics, Zhejiang University, Hangzhou,  Zhejiang 310027, China}
\address{Zhejiang Province Key Laboratory of Quantum Technology and Devices, Zhejiang University, Hangzhou 310027, China}

\date{\today}
\begin{abstract}
Geometrically frustrated interactions may render classical ground-states macroscopically degenerate. The connection between classical and quantum liquids and how the degeneracy is affected by quantum fluctuations is, however, not completely understood. 
We study a simple model of coupled quantum and classical degrees of freedom, the so-called Falicov-Kimball model, on a  triangular lattice and away from half-filling.
For weak interactions the phase diagram features a charge disordered state down to zero temperature. We provide compelling evidence that this phase is a liquid and show that it is divided by a crossover line that terminates in a quantum critical point.
Our results offer a new vantage point to address how quantum liquids can emerge from their classical counterparts. 
\end{abstract}
\maketitle

Liquids are characterized by the absence of long-range order. In exceptional cases, a classical liquid state may  persist down to zero temperature \cite{Wannier.50,Castelnovo2008}. 
The ensuing ground-state is macroscopically degenerate and characterized by a finite entropy.
Such a degeneracy in the energy landscape can result from competing interactions, geometric frustration, or near phase transitions where different states compete.  
The proliferation of low-energy states renders the system unstable towards the emergence of novel and often exotic groundstates.

An interesting and still open issue concerns the relation between classical and quantum liquids, {\itshape i.e.}, how quantum fluctuations affect the classical ground state manifold \cite{Rousochatzakis.18,Zhou.17}.
Another pertinent issue of practical relevance is the stability of such liquid phases with respect to, for example, the itinerary of the frustrated degrees of freedom.  
Any attempt of answering these questions is faced with the principal difficulty that quantum fluctuations of competing interactions, covering a wide energy range, need to be taken into account.

In this letter, we identify liquid phases which are driven by a coupling to quantum degrees of freedom. Depending on whether the quantum variables acquire a non-zero mass, the nature of the liquid phase changes.
To this end, we study the Falicov-Kimball model (FKM), a hybrid model comprised of itinerant fermions that interact arbitrarily strongly with localized charges,  on a triangular lattice. 
This model is  well suited to address some of the unresolved issues mentioned above and,  in addition, allows for an effective, numerically exact Monte-Carlo sampling of its partition function\cite{Maska2005, Antipov.16}. 


The FKM can be thought of as a special case of the Hubbard model with infinite mass imbalance between the two spin species thus rendering the dynamics of the heavier one classical. The position of these classical charges is annealed over an energy landscape defined by the itinerant degrees of freedom.
The model has been instrumental in benchmarking the standard approach to strongly correlated lattice models, {\itshape i.e.}, the dynamic mean field theory, and its extensions \cite{Metzner1989,Georges1996,Brandt1989,Freericks2003,Antipov2014,Rohringer.18}. More recently, it has attracted interest in the context of disorder-free many-body localization \cite{Antipov.16,Smith.17a,Smith.17b}. 
For bipartite lattices, several exact results have been established
\cite{Brandt1986,Kennedy1986}, including the existence of a charge density wave (CDW) at low temperature (T) for all interaction strengths.
The melting of the CDW state with increasing $T$ was observed for commensurate fillings \cite{Brandt1986,Kennedy1986,Maska2006,Zonda2009}.
 On the triangular lattice the FKM and  its extensions display a variety  of different ground-state phases \cite{Cencarikov.07, Yadav2010, Yadav2011, Yadav.11, Kumar.16}. For incommensurate fillings the FKM favors phase separation \cite{Maska2005}. Recently, it was demonstrated that the half-filled model on the square lattice is non-metallic at all non-vanishing values of the interaction strength $U$ and transitions from Anderson to Mott  insulators as $U$ is varied \cite{Antipov.16}. 


An effective model for the classical charges can be derived perturbatively at sufficiently small coupling $t/U$, where $t$ is the hopping strength of the itinerant electrons. At half-filling and for large $U/t$, the FKM is equivalent to the antiferromagnetic Ising model. Thus, on the square lattice, order ensues at sufficiently low T whereas the large-$U$ limit remains disordered for all T on the triangular lattice \cite{Wannier.50}. 
While it is an interesting question how this gets modified in the presence of quantum fluctuations \cite{Miguel.19}, the parent state at large $U$ is already a liquid. 
Here, however, we will in what follows demonstrate that a classical liquid state can result from an ordered phase, due to coupling to a quantum field. 


The Hamiltonian of the FKM is
\begin{equation}
H= -t \sum_{\left\langle ij \right\rangle} c^\dagger_{i} c_{j}  + U \sum_i c^\dagger_{i} c_{i}  n_{f,i} 
- \mu_c \sum_{i}   c^\dagger_{i} c_{i}   - \mu_f \sum_{i}  n_{f,i},
\end{equation} 
where $c^\dagger_{i}$ creates a $c$-electron and $n_{f,i}$, a conserved quantity,  counts the number of immobile classical charges on site $i$; and $\mu$ is the chemical potential of the system. $t$ will be used as a unit of energy ($t=1$). The summation $\sum_{\left\langle i,j \right\rangle}$ runs over all nearest-neighbor pairs on a triangular lattice with a volume $V=L^2$ ($L$ being the system's linear dimension) and periodic boundary conditions. 
As the collection of $n_{f,i}$ constitutes a set of conserved quantities, the partition function is given by a summation over non-interacting contributions for every configuration $n_f$ of f-charges and thus can be evaluated by an efficient Monte-Carlo sampling\cite{Maska2005, Antipov.16}. 
\begin{figure}[t!]
\centering
\includegraphics[width= 0.5 \textwidth]{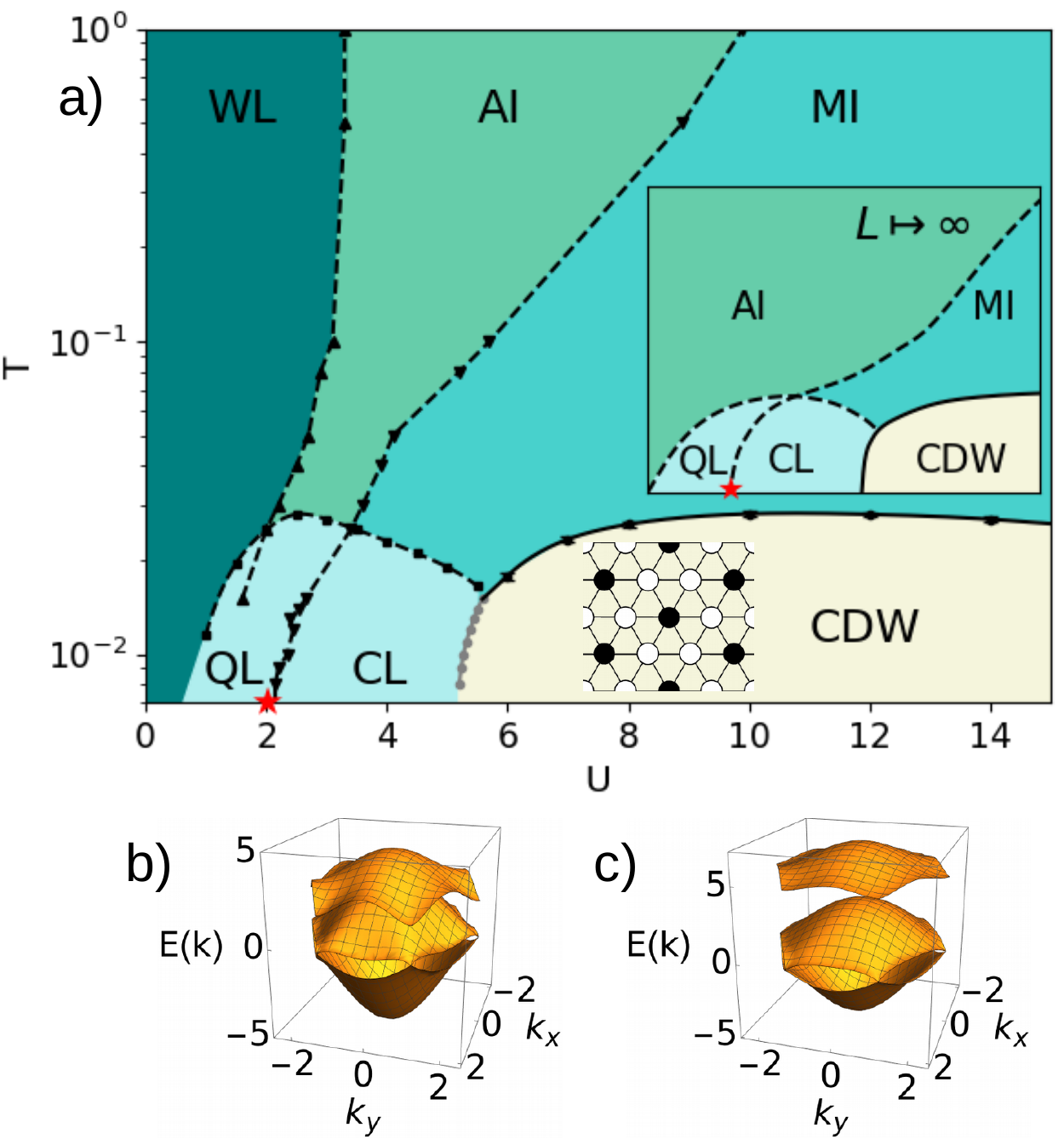} 
\caption{ (a) Phase diagram of the FK model on the triangular lattice in the $T-U$ plane for $x_f=1/3$ and $x_c=2/3$.  
The red star indicates the quantum phase transition between the liquid states. The inset depicts a sketch of the phases in the thermodynamic limit. 
Band structure assuming CDW order for $U=2$ (b) and $U=5$ (c).  
}
\label{phase_dia}
\end{figure}
An additional difficulty arises 
as the chemical potentials $\mu_{c}(T,U,L)$ and  $\mu_{f}(T,U,L)$ need to be self-consistently determined to ensure constant occupation as a function of $T,U,$ and $L$ \cite{SupMat}-S2.

In the following, we consider a macro-canonical ensemble with the chemical potentials  $\mu_c$ and $\mu_f$ determined such that $x_c=N_c/V = 2/3$ and $x_f = N_f/V = 1/3$, where $N_c=\av{\sum_{i}   c^\dagger_{i} c_{i} }$ and $N_f= \sum_{i}   \av{n_{f,i}} $, see however \cite{SupMat}-S6. 
\begin{figure}[t!]
   \begin{center}
     \includegraphics[width=.485\textwidth]{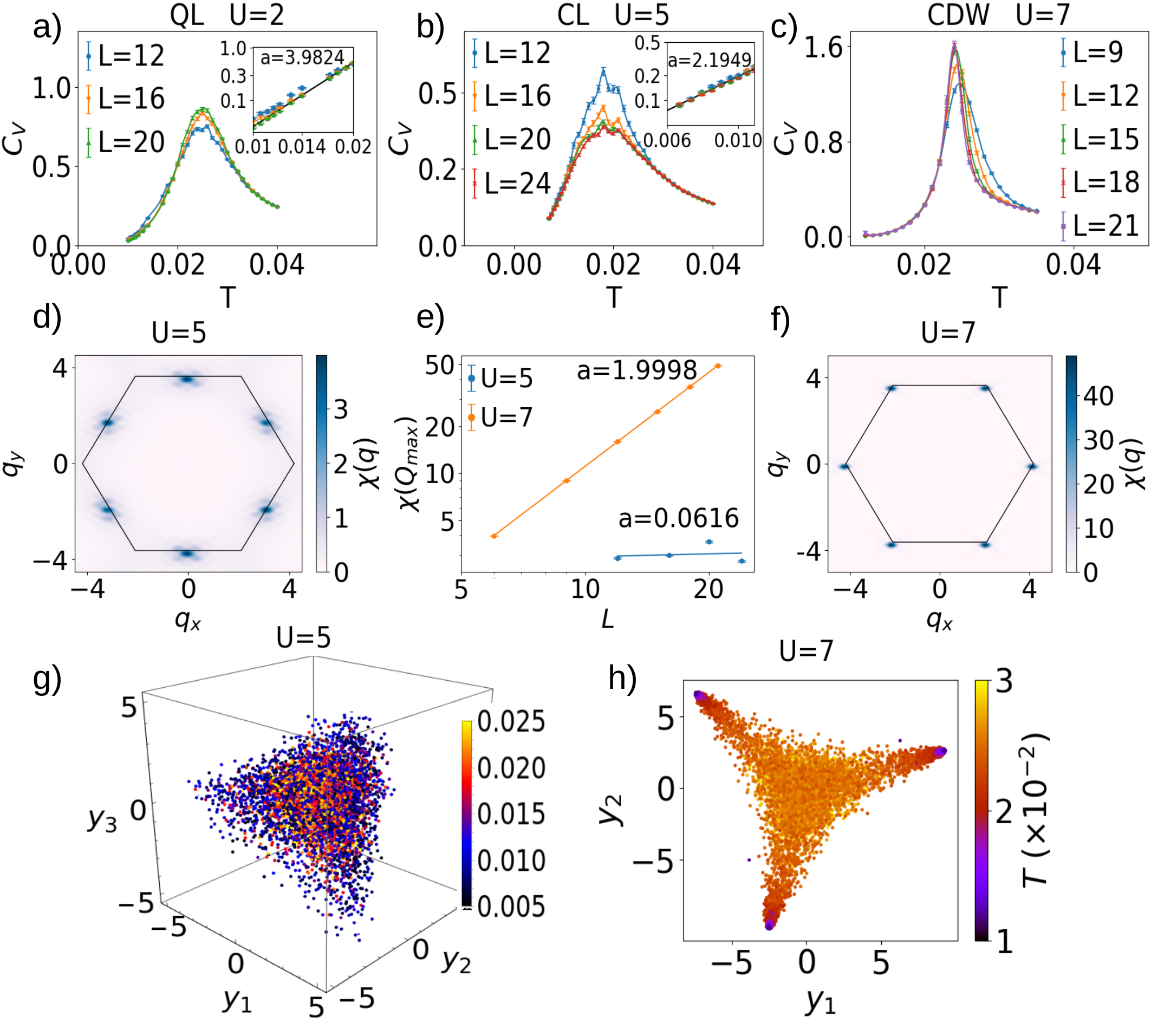} 
    \end{center} 
    \caption{\label{order_vs_disorder} Specific heat as a function of $T$ for $U=2$ (a), $U=5$ (b) and $U=7$ (c). Momentum resolved susceptibility of the f-charges $\chi(q)$ for $U=5$ (d) and $U=7$ (f). (e) finite size scaling of $\chi(Q_{\text{Max}})$ with $L$. PCA yielding (g) 3 most important components for $U=5$ and (h) 2 for $U=7$,  plotted for a range of $T$ for $L=20$ and $L=21$ respectively, \cite{SupMat}-S4. 
    }%
\end{figure}
Our findings are summarized in the phase diagram of
Fig.\ref{phase_dia}.
At large $U$ and sufficiently low $T$, ($T<T_c(U)$), the system develops a CDW. For large $T$ and/or $U$, the phase diagram resembles that of its counterpart on the square lattice.
While this could have been anticipated, there are important differences with regard to the type of order, see below.
The most surprising feature is the two regions at low $T$ and small $U$, labelled classical charge liquid (CL) and quantum charge liquid (QL), respectively, that remain disordered down to the lowest $T$. 
At large $U$, the $T$-driven phase transition is continuous and within the universality class of the 3-state  Potts model. 
For small $U$, our data point to a discontinuous transition between the CDW and the CL down to $T_c(U_c)=0$, corresponding to $U_c\simeq 5.2$.
For large $T$, a 'strange metal region' (WL) is found, where the electrons are weakly localized, followed by an Anderson insulating (AI) and finally a Mott insulting (MI) phase as $U$ increases which arises due to the  condition $x_c+x_f=1$,  generalizing the half-filling condition of the (spinfull) Hubbard model. The WL region  is only stabilized by the finite extent of the system and is expected to vanish in the thermodynamic limit \cite{Antipov.16}. AI is characterized by a finite density of states (DOS) at the Fermi level and a volume-independent inverse participation ratio (IPR). In the MI phase, the chemical potential lies within a $U$-dependent spectral gap.
This all closely resembles earlier findings for the FKM on a square lattice of reference \cite{Antipov.16}, where a full characterization of these phases, including a discussion of the optical conductivity, can be found. In what follows, we focus on the fundamentally new  features that emerge from the interplay of localized and itinerant degrees of freedom in a geometrically frustrated environment.

In the expansion in terms of $t/U$, higher order terms beyond the nearest-neighbour Ising-like interaction proportional to $t^2/U$ can be systematically derived \cite{Gruber1997}.
In next-to-leading order, i.e. $t^3/U^2$, it yields a term that couples the degrees of freedom on a triangular plaquette. The range of the effective interaction increases with powers of $t/U$, increasing the frustration which eventually leads to the melting of the order.  This  procedure is carried out in  \cite{SupMat}
up to order $t^4/U^3$. 

At large $U$, the effective term together with the $x_f=1/3$ restriction favors the existence of a low $T$ ordered phase possessing a 3-fold degeneracy. A possible order parameter for this phase is $\phi_{1/3}=\frac{3}{V}\sum_{r} e^{i\frac{2\pi}{3}(r_x-r_y)}n_{f,r}$ that equals $\phi_{1/3}=1, e^{i\frac{2\pi}{3}}$ or  $e^{i\frac{4\pi}{3}}$ depending on which of the three degenerate ground states is realized. 
One such configuration is depicted in the inset of Fig.\ref{phase_dia}, while the others can be obtained by a translation.

The order parameter symmetry implies that the associated finite-$T$ transition belongs to the two-dimensional 3-state Potts model universality class. 
We find for the correlation length exponent $\nu= 0.8031(114)$ ({\itshape cf.} $\nu_{3\text{Potts}}= 5/6$ \cite{Baxter}) and for $\gamma$, the exponent of the $T$ dependence of the susceptibility of the $f$-charges, $ \gamma = 1.4748(329)$ ({\itshape cf.}$\gamma_{3\text{Potts}}=13/9$ \cite{Baxter});
for details, see \cite{SupMat}.

The specific heat $C_v$ across the charge ordering transition for $U=7$ is shown in Fig. \ref{order_vs_disorder}-(c).
A high-$T$ local maximum (not shown in the figure) coincides with the $T$ scale where double-occupancy is sharply suppressed as $T$ is lowered. At lower $T$, the divergent $C_v$, confirmed by the finite-size scaling, corresponds to the transition into the CDW state.  
The static susceptibility, $\chi(\omega\rightarrow 0,q,T)$ is depicted in Fig.\ref{order_vs_disorder}-(f). It shows a maximum at the propagation vector $Q_{\text{CDW}}$ of the CDW, {\itshape i.e.}, $q=Q_{\text{CDW}}= 2\pi/3\{1,\sqrt{3}\}$. Fig.\ref{order_vs_disorder}-(e) (orange line) shows that, for $T<T_{c}$,  $\chi(\omega\rightarrow 0,\boldsymbol{ Q}_{\text{CDW}},T) \propto L^2$, in line with the existence of long-range order.
The  reconstructed band structure of the $c$-electrons within this symmetry broken phase at $T=0$ is
given in Fig.\ref{phase_dia}-(c); $\mu_c$ lies in the $U$-dependent bandgap.    
Assuming that the ordered phase persists as a function of $U$ one expects an indirect closing of the bandgap for $U=3$.  Fig.\ref{phase_dia}-(b) shows the band structure for $U<3$. The charge order, however, vanishes at $U_c >3$, see Fig.\ref{phase_dia}-(a).

We now address the region $U<U_c$. Fig.\ref{order_vs_disorder}-(a) and (b) show  $C_v(T)$ in this region for a representative value of $U$ within the QL and CL respectively. The high-$T$ features are similar to those found for $U>U_{c}$.
In contrast to the $U>U_c$ case, $C_v(T,L)$ remains non-singular for $L\rightarrow \infty$, indicating that the transition has been replaced by a crossover. 
The insets of Figs.\ref{order_vs_disorder}-(a) and (b) depict $C_v(T)$ on a logarithmic scale and indicate that the behavior is compatible with power-law scaling in $T$ implying gapless excitations in the system.
The fact that both power-laws are distinct highlights that there are indeed two different $T=0$ phases. Moreover, neither is compatible with Fermi-liquid behavior, {\it i.e.}, with $C_v(T)\propto T$. 
Within these phases, $\chi(0,\boldsymbol{q},T)$ is no longer maximal for $\boldsymbol{q}=\boldsymbol{Q}_{\text{CDW}}$ but rather for $\boldsymbol{q}=\boldsymbol{ Q}_{\text{Max}}=  \pi\{1,1/\sqrt{3}\} $, as shown in Fig.\ref{order_vs_disorder}-(d) for CL (similar for QL). Interestingly, $\boldsymbol{ Q}_{\text{Max}}$ corresponds to the wave vector of a CDW expected for filling fractions $x_f=1/4$ and $x_c=3/4$, see Fig.\ref{order_vs_disorder}-(d). 
However, the order parameter $\phi_{1/4}$ of this phase, explicitly given in \cite{SupMat}, vanishes (see below).  
The scaling of $\chi(0,\boldsymbol{Q}_{\text{Max}},T)$ with $L$ shown in Fig.\ref{order_vs_disorder}-(e) (blue line) is  $\chi(\omega\rightarrow 0,\boldsymbol{ Q}_{\text{Max}},T) \propto L^a$ with $a\simeq 0.0616 $, which indicates that the CL region is incompatible with the existence of long-range order of that type. 
To further substantiate the characterization of the liquid region, we turn to a principal component analysis (PCA)\cite{Jolliffe.14} of the charge excitations in CL and CDW. 
 This method allows for a dimensional reduction when visualizing multivariate data \cite{Wang.16}. Figs.\ref{order_vs_disorder}-(g) and (h) show the projection of different thermalized configurations onto the three principal components obtained by a PCA analysis including uncorrelated configurations at different $T$, see \cite{SupMat}-S4.
 In the ordered phase, low-$T$ configurations cluster around one of the three ground-states, see Fig.\ref{order_vs_disorder}-(h) which correspond to two principal components. 
 Within the CL region, however, configurations cluster on a four-fold symmetric structure corresponding to three principal components.
 Note, however, that this does not imply the existence of a long-range ordered four-fold state, which is incompatible with the $1/3$ filling, unless phase separation occurs. 
A vanishing order parameter and associated susceptibility $\chi(0,\boldsymbol{ Q}_{\text{Max}},T)=\chi(L)$ as L$\rightarrow \infty$ is, however,  incompatible with phase separation, see Fig.\ref{order_vs_disorder}-(e) \cite{SupMat}.
\begin{figure}
    \begin{center}
    \includegraphics[width=0.95\linewidth]{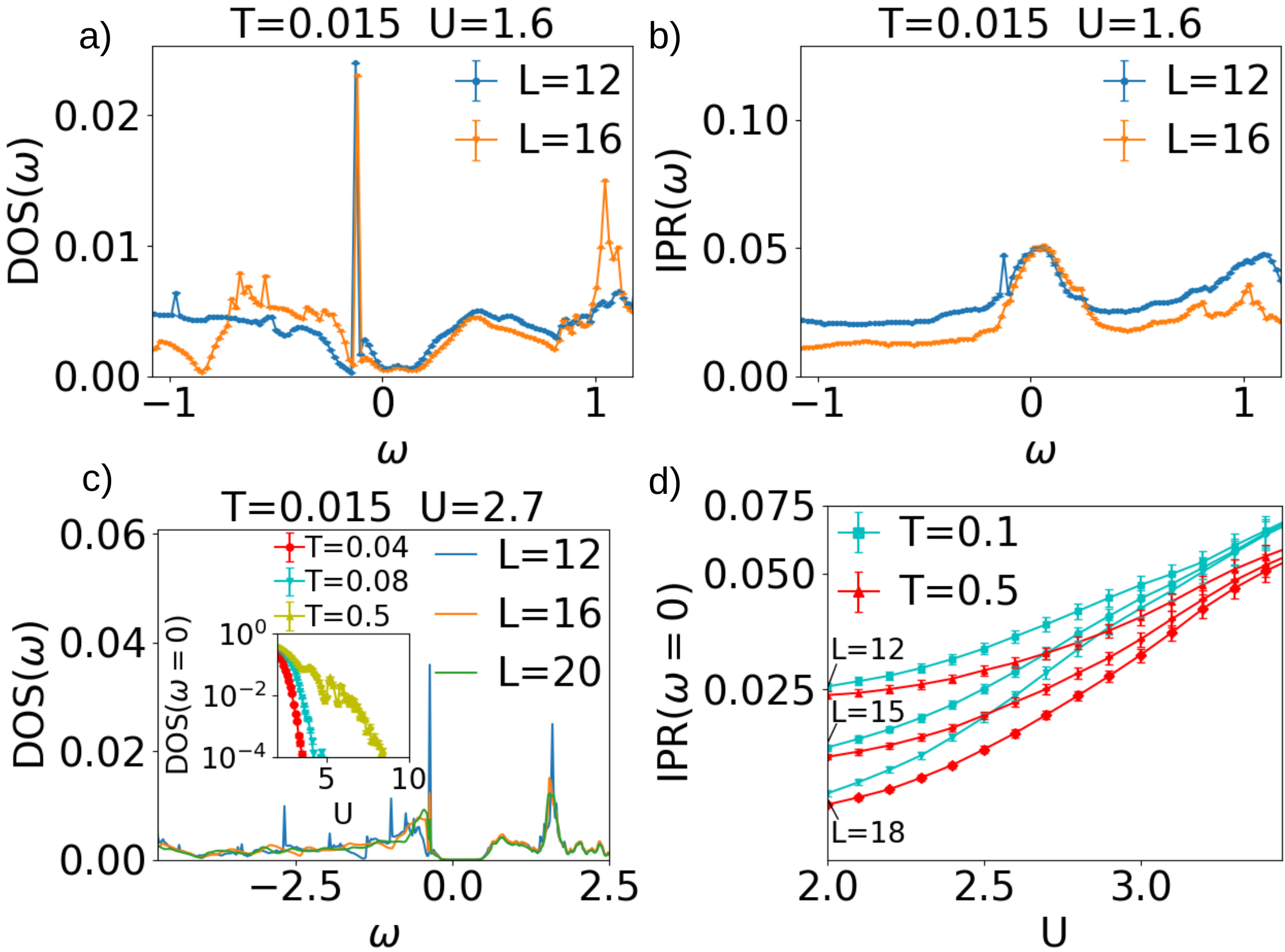}
    \end{center}\vspace{-1em}
    \caption{\label{c_properties}$c$-electron properties:  DOS for $U=1.6$ (a) and $U=2.7$ (c) for $T=0.015$. Inset of (c) - $\text{DOS}(\omega=0)$ vs $U$ for different $T$.   
    IPR as a function of energy for different system sizes (b).  
    (d) Scaling of the  $\text{IPR}(\omega=0)$ as a function of $U$. Data obtained from  simulations with 100000 Monte-Carlo steps with correlation times of at most 20 steps.
    }%
\end{figure}

 Another possibility in line with the power-law behavior in $\chi(L)$ and $C_v(T)$ for $T\to0$ is the occurrence of a phase transition, at a $T$ near the low-$T$ peak in the specific heat, to a KT-like phase, lacking long-range order. Indeed,  KT transitions can occur for clock-models with $Z_q$ symmetry \cite{Elitzur.79}. While our numerical results alone cannot exclude this scenario, this only arises for clock-models with $q>4$ \cite{Elitzur.79, Rujan.81, Baek.10}. Moreover the systems seems to retain its full symmetry, isomorphic to the permutation group and not just a $Z_4$ subgroup. 
  
The only viable alternative for the low $T$, low $U$ regions of the phase diagram is the existence of charge liquid states, connected by a smooth crossover to the charge-disordered state at high $T$. The observed behavior in $C_v$ vs $L$ is indeed in line with such a crossover, see Fig.\ref{order_vs_disorder}-(a,b) as is its  $T\rightarrow 0$ behavior, as discussed above.
 
In order to understand the difference between CL and QL we study the properties of the $c$-electrons through their DOS and IPR.
Fig.\ref{phase_dia} shows the liquid region is intersected by the AI-MI crossover line defined by the vanishing of the DOS at zero energy. 
At $T=0$, this crossover line should terminate in a continuous Mott transition separating two phases where the $c$-electrons pass from being gapless to being gapped as $U$ increases. 
Fig.\ref{c_properties}-(a) and  Fig.\ref{c_properties}-(c), show the DOS as a function of energy for two values of $U$; one below (a) and the other (c) for $U$ above the AI-MI crossover line separating the QL and CL regions. 
Fig.\ref{c_properties}-(c) shows that in the MI side there is a region around $\omega = 0 $ without $c$-electron states in  contrast to  the finite DOS around $\omega = 0 $ seen in Fig.\ref{c_properties}-(a). The inset of Fig.\ref{c_properties}-(c) depicts the behaviour of the $\text{DOS}(\omega=0)$ vs $U$ for different $T$ showing a sharp drop. This sharp drop is used to determine the crossover line. 
 
Fig. \ref{c_properties}(b) shows an example of the $\text{IPR}(\omega)$ in the vicinity of the Fermi energy and its scaling with system size for a case where $\text{IPR}(\omega=0)$ converges to a finite value as a function of $L$.
Fig.\ref{c_properties}-(d) depicts the scaling of $\text{IPR}(\omega=0)$ as a function of system size. 
The crossover line separating the WL from the AI region signals the localization-delocalization transition of the $c$-electrons and is obtained from the scaling behavior of $\text{IPR}(\omega=0)$ with volume. 
This $\text{IPR}(\omega=0)$ line can be continued into the QL region, see Fig.\ref{phase_dia}. 

The WL is expected to vanish as the thermodynamic limit is taken \cite{Antipov.16}. However, there is the interesting possibility that the fate of the delocalized region as the thermodynamic limit is taken, is different from that of WL, as the sampled disorder configurations in the CL region are highly correlated. This would allow for an itinerant $c$ electron phase in the thermodynamic limit. Our data, however,  do not permit us to discriminate between these scenarios due to dominating finite-size effects.

\begin{figure}
    \begin{center}
   \includegraphics[width=0.5\textwidth]{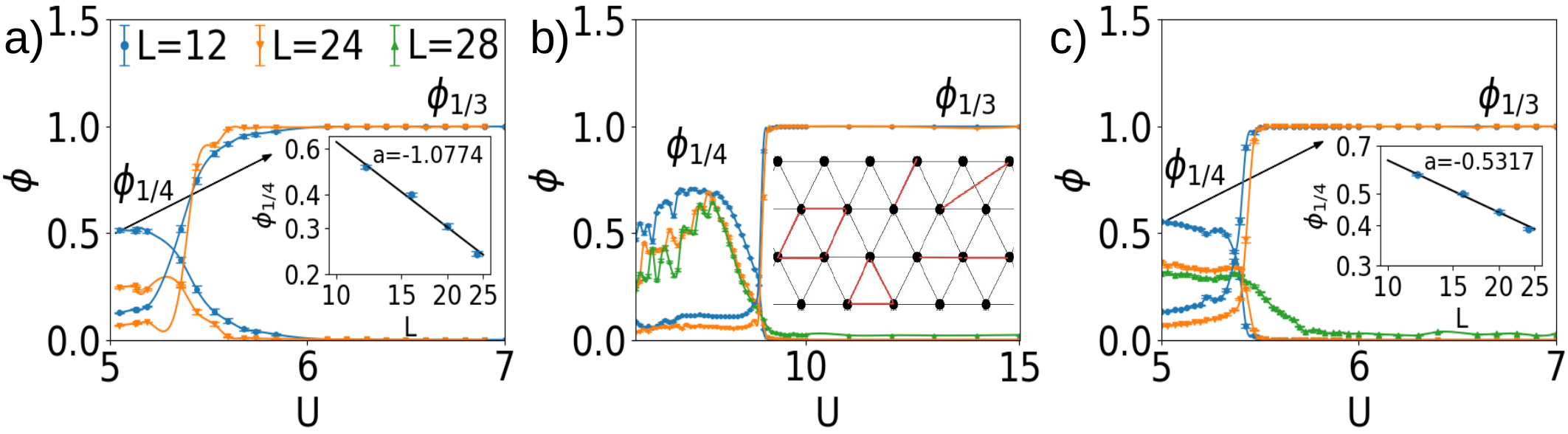}
    \end{center}\vspace{-1em}
    \caption{Order parameters $\phi_{1/3}$ and $\phi_{1/4}$ for (a) the FK model,  (b) the effective classical model truncated to 4th order,  (c) an effective variatonal model , with couplings  determined by a linear regression.  The inset of (a) and (c) show  the scaling of the $\phi_{1/4}$ with system sizes for $U=5$, respectively for the FK and variatonal models.  
    The inset in (b) depicts the schematic couplings between sites obtained up to 4th order.  
     \label{fig:classical}
    }%
\end{figure}

Fig. \ref{fig:classical}- (a) shows the order parameters of $\phi_{1/3}$ and $\phi_{1/4}$ of the (full) FK model. 
The inset shows that the amplitude of $\phi_{1/4}$ vanishes in the thermodynamic limit as $L^a$ with $a\simeq -1$. 

Within our numerical precision, the nature of the CL-CDW transition in  Fig. \ref{fig:classical}-(a) is compatible with a discontinuous  transition with double peaked distribution of the order parameter and energy. The  Binder cumulant, however, remains positive near the transition (see \cite{SupMat}). 
In order to avoid spurious finite size effects we only consider 
$L=12,24$, which are commensurate with both the CDW and the incipient $1/4$-filling ordered background.

To further analyze the origin of the CL phase  we turn to a study of the effective classical model, see \cite{SupMat}, using a Monte Carlo algorithm. 
Fig. \ref{fig:classical}-(b) depicts the same quantities but for the effective classical model obtained by truncating the expansion to fourth order and where the effective couplings are replaced by their  $T=0$ limits. The  inset of \ref{fig:classical}-(b) schematically depicts the different coupling terms of the effective Hamiltonian. In contrast with the FK model the truncated one exhibits a phase separated state at small $U$ characterized by a non-vanishing value of $\phi_{1/4}$.    
Fig. \ref{fig:classical}-(c) shows $\phi_{1/3}$ and $\phi_{1/4}$ for an effective model with the same type of interactions as (b) but with the coupling constants determined by linear regressions \cite{Liu.17}, explicitly given in  \cite{SupMat}. 
The inset demonstrates that $\phi_{1/4}$ indeed vanishes albeit with a different power of the linear system size $a\sim - 1/2$.

In the QL, the gapless quantum degrees of freedom induce long-range interactions among the classical charges. In contrast, for the CL, we expect that an effective Hamiltonian exists in terms of short-ranged classical charges.
Up to fourth order, however, neither the truncated model nor the variational one seem to capture the properties of the CL phase. 
Apparently, higher order terms are necessary to fully capture the properties of this phase. 
We thus have obtained an unexpectedly rich phase diagram of the $1/3$-filled FKM on the triangular lattice:
for intermediate-to-large coupling, the $T$-driven phase transition from the charge-ordered to the disordered phase belongs to the 3-state Potts model universality class; more importantly, for weak coupling and low $T$, we show the existence of a charge liquid region divided by a cross-over line that terminates in a QCP at $U_{\mbox{\tiny QCP}}$.
The classical liquid, which ensues for $U>U_{\mbox{\tiny QCP}}$, is expected to be captured by an effective, finite-ranged  classical model. However, we were not able to obtain such a model with terms  up to the fourth order  in terms of $t/U$.
Approaching the QCP from within the CL, the gap of the quantum degrees of freedom vanishes at $U_{\mbox{\tiny QCP}}$. This phase transition is reflected in the behavior of the specific heat which changes from $C_v \propto T^2 $  to $C_v \propto T^4$ as   $U_{\mbox{\tiny QCP}}$ is crossed.
 To better understand the differences between the full and the truncated classical model it will be instructive to
 analyze the effect of higher orders systematically \cite{Yang_2012}. 
 
Our results show how quantum liquids 
can emerge from their classical counterparts.
They also elucidate how classical liquids form through melting of phase separated states in the presence of frustrated interactions.
Addressing the fate of the charge liquids in the presence of non-vanishing hybridization between the localized charges and the conduction electrons is an interesting open question with immediate experimental relevance \cite{Nakatsuji.06, Tokiwa.14}. Understanding the fate of the CL phase as the FKM approaches the Hubbard model may help unraveling the phase diagram of the Hubbard model on the triangular lattice \cite{GangLi.14}.


\begin{acknowledgments}
\vspace{10pt}
We gratefully acknowledge  helpful discussions with  A.\ Antipov, R.\ Mondaini, A.\ Sandvik and D.\ Vollhardt.
Computations were performed on the Tianhe-2JK cluster at the Beijing Computational Science Research Center (CSRC) and on the  Baltasar-Sete-S\'{o}is cluster, supported by V. Cardoso's H2020 ERC Consolidator Grant no. MaGRaTh-646597, computer assistance was provided by CSRC and CENTRA/IST. 
M.\ M.\ Oliveira acknowledges partial support by the FCT through IF/00347/2014/CP1214/CT0002 and SFRH/BD/137446/2018.
P. Ribeiro acknowledges support by FCT through the Investigador FCT contract IF/00347/2014 and Grant No. UID/CTM/04540/2013.
S.\ Kirchner acknowledges support by  the National Science Foundation of China, grant No.\ 11474250 and No.\ 11774307 and the National Key R\&D Program of the MOST of China, Grant No.\ 2016YFA0300202. 
\end{acknowledgments}

\bibliographystyle{apsrev4-1}
\bibliography{FKreferences}

\newpage 

\begin{widetext}
\begin{center}
\textbf{\large{}\textemdash{} Supplemental Material \textemdash{}}
\par\end{center}{\large \par}
\begin{center}
\textbf{\large{}Classical liquid induced by quantum fluctuations}
\par\end{center}{\large \par}
\begin{center}
\textbf{Miguel M. Oliveira$^{(a)}$, Pedro Ribeiro$^{(a,b)}$, and Stefan Kirchner$^{(c,d)}$}\\
$^{(a)}$CeFEMA, Instituto Superior T\'ecnico, Universidade de Lisboa Av. Rovisco
Pais, 1049-001 Lisboa, Portugal\\
$^{(b)}$Beijing Computational Science Research Center, Beijing 100193, China\\
$^{(c)}$Zhejiang Institute of Modern Physics, Zhejiang University, Hangzhou,  Zhejiang 310027, China\\
$^{(d)}$Zhejiang Province Key Laboratory of Quantum Technology and Devices, Zhejiang University, Hangzhou 310027, China
\end{center}
\begin{description}
\item [{Summary}] Below we provide additional technical details and further numerical results supplementing the conclusions from the main text.
\end{description}
\end{widetext}

\beginsupplement 

\section{Monte Carlo Algorithm} 
\label{sec:MCA}

The partition function of the model can be written as  
\begin{equation}
Z = \sum_{\ \{ n_f \} } e^{-\beta F(n_f)} ,
\end{equation} 
with 
\begin{equation}
F(n_f)= -\frac{1}{\beta} \sum_{\nu} \log \left[ 1 + e^{-\beta \left[ \epsilon_{\nu}(n_f) - \mu_c \right]} \right] - \mu_f N_f.
\end{equation} 
being the free energy of a given $n_f$ electron configuration, and $\epsilon_{\nu}(n_f)$ the eigenvalues of the single-particle Hamiltonian for the c-electrons. 

The mean value of any observable $O$ depending only on the classical charges, is thus computed as 
\begin{equation}
\langle O \rangle = \sum_{\ \{ n_f \} } O(n_f) e^{-\beta F(n_f)}. 
\end{equation}

\section{Determination of the chemical potentials} 

The determintion of the chemical potentials $\mu_c(T,L,U)$ and $\mu_f(T,L,U)$ associated with given fillings has to be performed with sufficient precision before observables can be accessed reliably. 
This procedure was not needed for  the half-filled FKM on a square lattice where particle-hole symmetry was used to fix the chemical potentials.  

This is accomplished via the Newton-Raphson algorithm, to find the roots, $\bold{f(\bold{x})}=\bold{0}$, of a function $\bold{f(\bold{x})}=(f_1(\bold{x}), f_2(\bold{x}), \cdots, f_k(\bold{x}))$ by iterating the set of equations
\begin{equation}
J_f(\bold{x}_n)(\bold{x}_{n+1}-\bold{x}_n)=-\bold{f}(\bold{x}_n) ,
\end{equation}
where $J_f(\bold{x})$ is the Jacobian matrix.
This method was used to fix the filling fractions of both electron species, $n_c=N_c/V$ and $n_f=N_f/V$ defining
\begin{equation}
\begin{cases}
f_1(\mu_f,\mu_c)=n_f(\mu_f,\mu_c)-1/3=0 \\[0.5em]
f_2(\mu_f,\mu_c)=n_c(\mu_f,\mu_c)-2/3=0
\end{cases}
.
\end{equation}
For each iteration we performed a Monte-Carlo simulation using the current iteration for the chemical potentials. 
During the simulation we gather data to compute  the fillings of both species as well as the Jacobian matrix. 
The results of the simulation were then used to update the chemical potentials according to
\begin{equation}
 \begin{pmatrix} \frac{\partial f_1}{\partial\mu_f} & \frac{\partial f_1}{\partial\mu_c} \\[0.8em] \frac{\partial f_2}{\partial\mu_f} & \frac{\partial f_2}{\partial\mu_c} \end{pmatrix}   \begin{pmatrix} \mu_{f,n+1}-\mu_{f,n} \\[0.3em] \mu_{c,n+1}-\mu_{c,n}  \end{pmatrix} =  -\begin{pmatrix} f_1(\mu_{f,n};\mu_{c,n}) \\[0.3em] f_2(\mu_{f,n};\mu_{c,n}) \end{pmatrix} ,
\end{equation}
to obtain $(\mu_{f,n+1},\mu_{c,n+1})$. The procedure was repeated until the fillings converged to the desired accuracy. 
Determining $\mu_f$ and $\mu_c$  turned out to be the computationally most expensive part of the numerical algorithm, especially at low $U$ and $T$ and thus became the main limiting factor in our numerical analysis.




\section{Details of the computation of DOS and IPR} 


For fixed $f$-electron configuration the density of states reads 
\begin{equation}
\text{DOS}_{n_f}(\omega) = \sum_\nu \delta\left[\omega-\epsilon_\nu(n_f) \right].
\end{equation}
The FK DOS is obtained by averaging over all possible configurations
\begin{equation}
\text{DOS}(\omega) = \frac{1}{Z} \sum_{\{n_f\}} \text{DOS}_{n_f}(\omega) e^{-\beta F(n_f)}. 
\end{equation}
The IPR for a specific state $|\psi_\nu(n_f)\rangle$ is defined as 
\begin{equation}
\text{IPR}_{|\psi_\nu(n_f)\rangle}= \frac{\sum_r |\langle r | \psi_\nu(n_f) \rangle|^4 }{\sum_r |\langle r | \psi_\nu(n_f) \rangle|^2},
\end{equation}
and thus, the energy resolved IPR becomes 
\begin{equation}
\text{IPR}_{n_f}(\omega) = \frac{\sum_\nu \text{IPR}_{|\psi_\nu(n_f)\rangle} \delta\left[\omega-\epsilon_\nu(n_f) \right]}{\sum_\nu \delta\left[\omega-\epsilon_\nu(n_f) \right]},
\end{equation}
which after being averaged over configurations results in
\begin{equation}
\text{IPR}(\omega) = \frac{1}{Z} \sum_{\{n_f\}} \text{IPR}_{n_f}(\omega) e^{-\beta F(n_f)}. 
\end{equation}

For the numerical evaluation we regularize the delta-functions either by assigning a finite width or by integrating over a small energy window. For all results presented in this work  the latter regularization procedure was used. The DOS of energy window $\Delta_i$ is given by
\begin{equation}
DOS_i= \frac{C_i}{N L^2}, 
\end{equation}
where $C_i$ is the number of eigenvalues of all sampled configurations within $\Delta_i$. $N$ is the number of samples used and $L$ is the linear system size.  Similarly, 
\begin{equation}
\text{IPR}_i= \frac{\sum_{j (\varepsilon_j \in \Delta_i )} \text{IPR}_{|\psi_j\rangle} }{C_i}, 
\end{equation}
where $j$ labels the eigenstates with eigenenergy $\varepsilon_j$. 

The errorbars for the DOS within each window are estimated as the standard deviation of the number of counts assuming a Poison distribution, i.e. $\sqrt{C_i}$. 
The error for the IPR follows through error propagation.

In the inset of Fig.3-(c) of the main text, we estimate the DOS at the Fermi level by a finite size scaling procedure that consists of decreasing the size of the energy window proportionally to the volume. 
The results are extrapolated to the thermodynamic limit to deduce if the system is gapped or not.

\section{Details of the computation of PCA} 

In the following, we provide further details about the principal component analysis (PCA) presented in Figs.2 -(g) and -(h) of the main text. The procedure follows closely that described in Ref.\cite{Wang.16}.

PCA is often employed for performing dimensional reduction of multivariate data. 
It consists of finding the directions within a data set for which variations are considerably larger and discarding those where variations are small.
This method allows to visualize the data set in a reduced dimensional space while minimizing relevant information loss.

We adopted the following procedure: 

- An $N\times L^2$ matrix, $X$, represents the $f$-electron configurations, such that  $X_{ir}$ corresponds to the $f$-electron occupation $n_{f,r}$ of sample $i=1,...,N$ which runs over uncorrelated configurations within a given range of temperatures. 

-  The covariance matrix estimator is computed explicitly by
\begin{equation}
(\tilde X^T \tilde X)_{rr'} = \langle (n_{f,r}-\langle n_{f,r} \rangle_{\text{conf}}) (n_{f,r'}-\langle n_{f,r'} \rangle_{\text{conf}} ) \rangle_{\text{conf}} ,  
\end{equation}
where $\tilde X_{ir} = X_{ir} - \frac{1}{N}\sum_i X_{ir}$, and the $\langle n_{f,r'} \rangle_{\text{conf}}$ corresponds to the averaging over all configurations $i=1...,N$; 

-  $\tilde X^T \tilde X$ is diagonlized and the data set is projected along the eigenvectors corresponding to the dominating eigenvalues;  

Figs.2-(g) and -(h) are computed following this procedure. 
We confirmed that for Figs.2-(h) two eigenvalues are dominating, reflecting the nature of the order parameter. Likewise, for the case in Figs.2-(g) three eigenvalues dominate the spectrum. 
 
For the CDW, Fig.2-(h), the low temperature data cluster around 3 vertices corresponding to the tree symmetry broken ground-states. For sufficiently high temperatures the data spreads out reflecting the absence of order.

\begin{figure}[t!]
\centering
\includegraphics[width= 0.5 \textwidth]{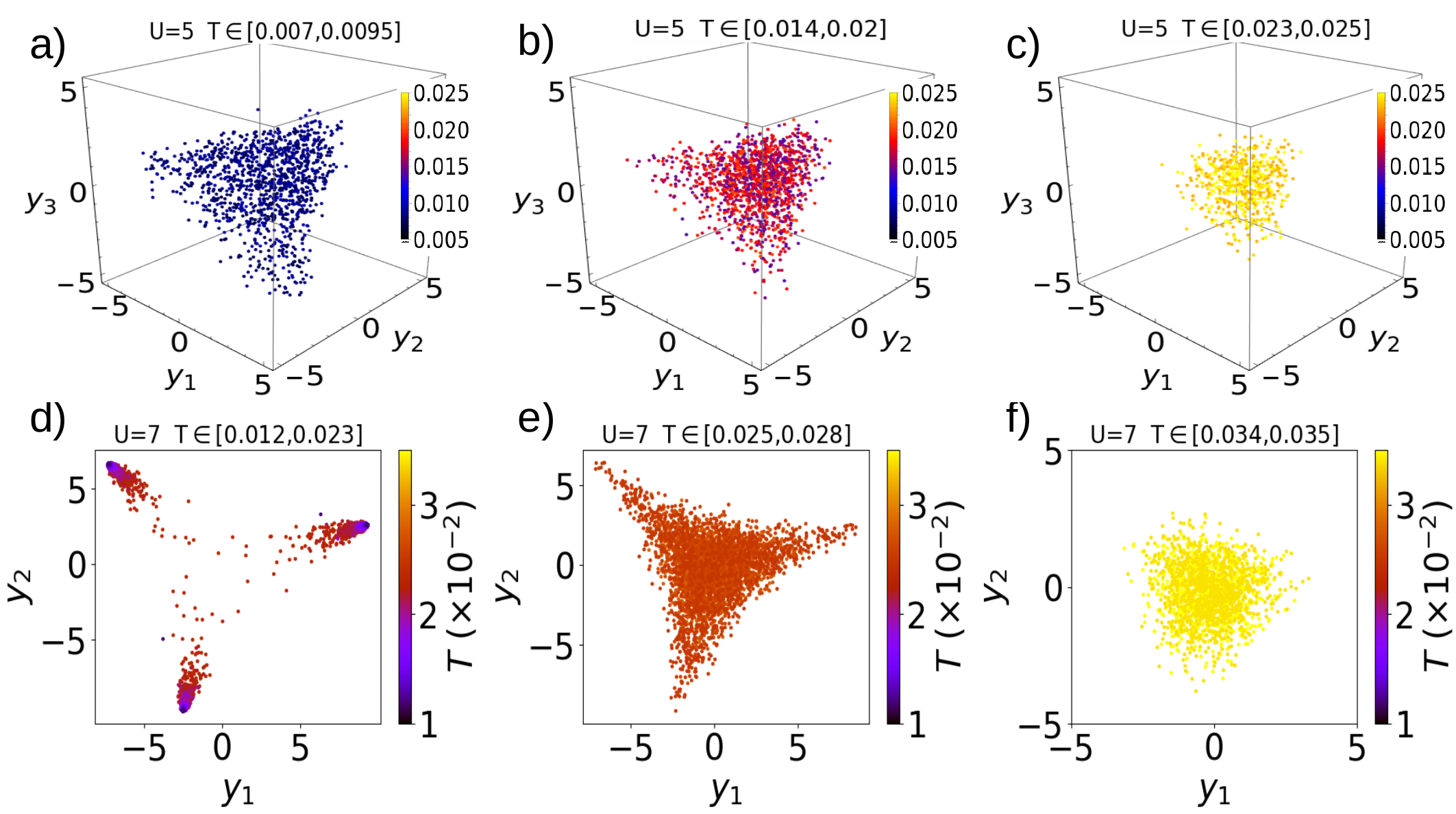} 
\caption{ 
PCA computed for the two values of $U$ which are for low $T$ part of the CL ($U=5$) and CDW ($U=7$) region respectively. The data is the same as in Fig.2-(g) and (h) of the main text but  here is split into smaller temperature windows for illustration purposes. The $U=5$ ($U=7$) data was computed for $L=20$ ($L=21$) and corresponds to $200$ ($1000$) uncorrelated configurations per temperature with 24 temperature values split into the different windows.
}
\label{fig:pca}
\end{figure}

For the CL region, Fig.2-(g), a tetrahedral structure emerges reflecting the symmetry of the putative order. 
However, the clustering is much less pronounced than for the CDW which is in line with the absence of an ordered state at the thermodynamic limit.  
Fig.\ref{fig:pca} shows the PCA computed for the two values of $U$ which are at low $T$ part of the CL and CDW region. 
The PCA around $T=0.035$  and $U=7$   is equivalent to the one around $T=0.025$  and $U=5$. Apparent differences are due to using  the two (three) most important components for $U=7$ ($U=5$). The most important components in each case where obtained from the analysis of $\tilde X^T \tilde X$ corresponding to eigenvalues with magnitudes  much larger than the average. 

\section{Calculation of the $f$-electron charge susceptibility}

Since the $f$-charges are conserved, their charge susceptibility in momentum space is  non-vanishing only for zero frequency, $\chi(\omega\to 0, \boldsymbol{q}, T)$. This quantity is shown in Figs.2-(d) and -(f), and is computed as
\begin{equation}
\chi(0,\boldsymbol{q},T)= \sum_{ij} \left[   \langle n_{f,i} n_{f,j} \rangle - \langle n_{f,i} \rangle \langle n_{f,j} \rangle \right]  e^{i \boldsymbol{q} \cdot (\boldsymbol{r}_i-\boldsymbol{r}_j)}. 
\end{equation}

\section{Canonical ensemble for the f-electrons}

\begin{figure}[t!]
\centering
\includegraphics[width= 0.5 \textwidth]{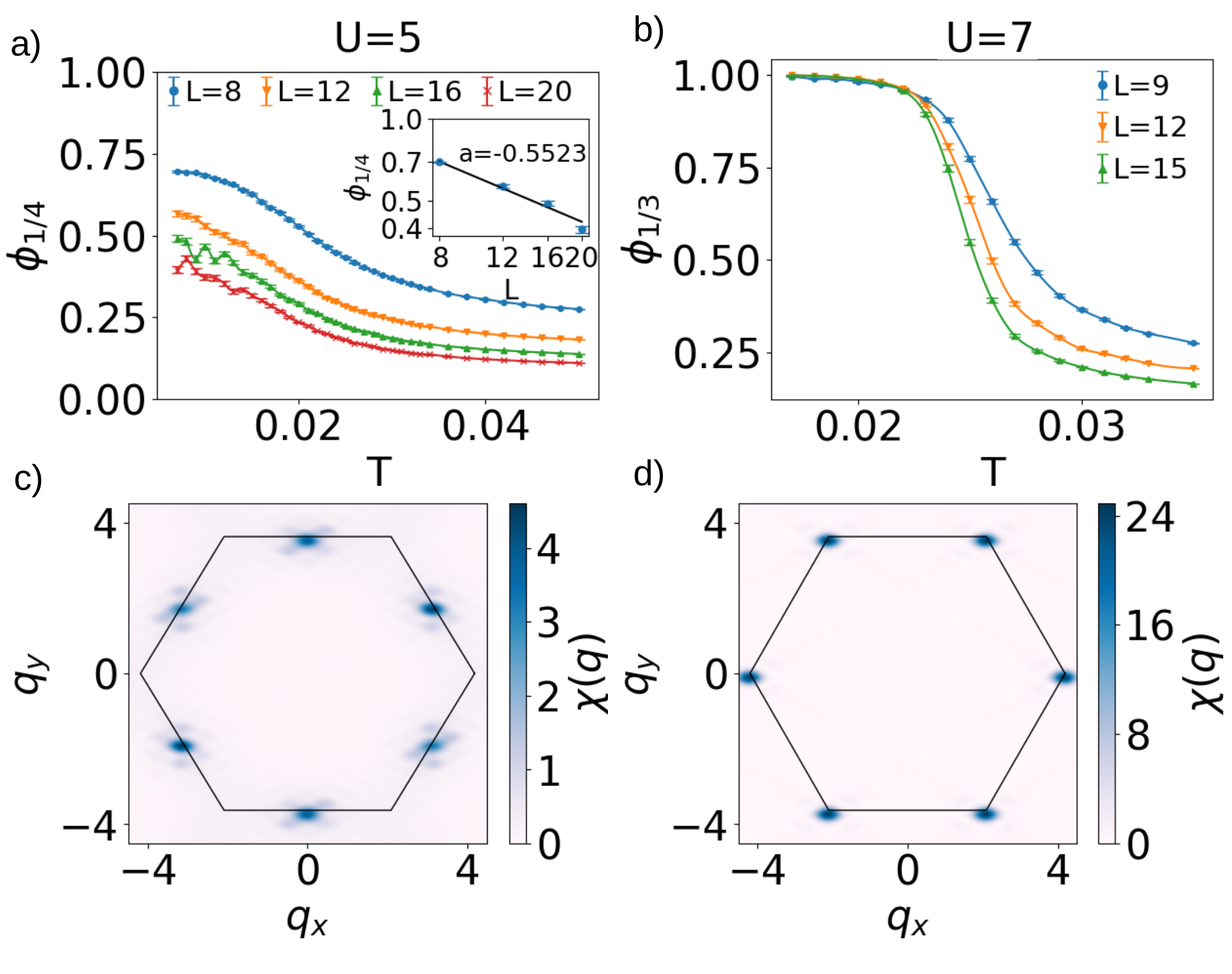} 
\caption{ 
MC simulation using the canonical ensemble for the $f$-electrons computed for two different values of $U$, which for low $T$ belong to the CL and CDW phases, respectively. (a) Order parameter $\phi_{1/4}$ for $U=5$. 
(b) Order parameter $\phi_{1/3}$ for $U=7$.
(c) and (d) - Momentum resolved susceptibility of the $f$-charges for $U=5$ and $U=7$ respectively.
}
\label{fig:canonical}
\end{figure}

Our results are based on a MC algorithm using the grand-canonical ensemble described in Sec.\,\ref{sec:MCA}. In principle, a similar procedure can be used for fixed $f$ particle number.  While the canonical procedure has the advantage of not having to solve for the associated chemical potential, it turns out that the grand-canonical algorithm is characterized by shorter Monte Carlo autocorrelation times and higher acceptance rates. This observation seems to be true more generally \cite{Zhenjiu.17}. In this section we explicitly demonstrate that both approaches yield equivalent results in the thermodynamic limit.
Fig.\ref{fig:canonical} shows the order parameter and the momentum resolved susceptibility for $U=5$ and $U=7$, for which the low temperature phases are respectively CL and CDW. 
For $U=5$ the order parameter  vanishes with system size as $\phi_{1/4} \propto V^{-1/2}$. 
Note that this scaling differs from the $\phi_{1/4} \propto V^{-1}$ obtained in the grand-canonical case. Similar differences in the finite-size scaling behaviour for canonical and grand-canonical ensembles were reported in Ref. \cite{Zhenjiu.17}. 
Apart from this difference all the computed quantities are equivalent. In particular, both the specific heat at fixed particle number, $C_v^{N_f}$, and  at fixed chemical potential, $C_v$, shown in the main text,  display power law behavior in the CL region and thus indicate the presence of gapless excitations.

\section{Potts transition}

\begin{figure}[t!]
\centering
\includegraphics[width= 0.5 \textwidth]{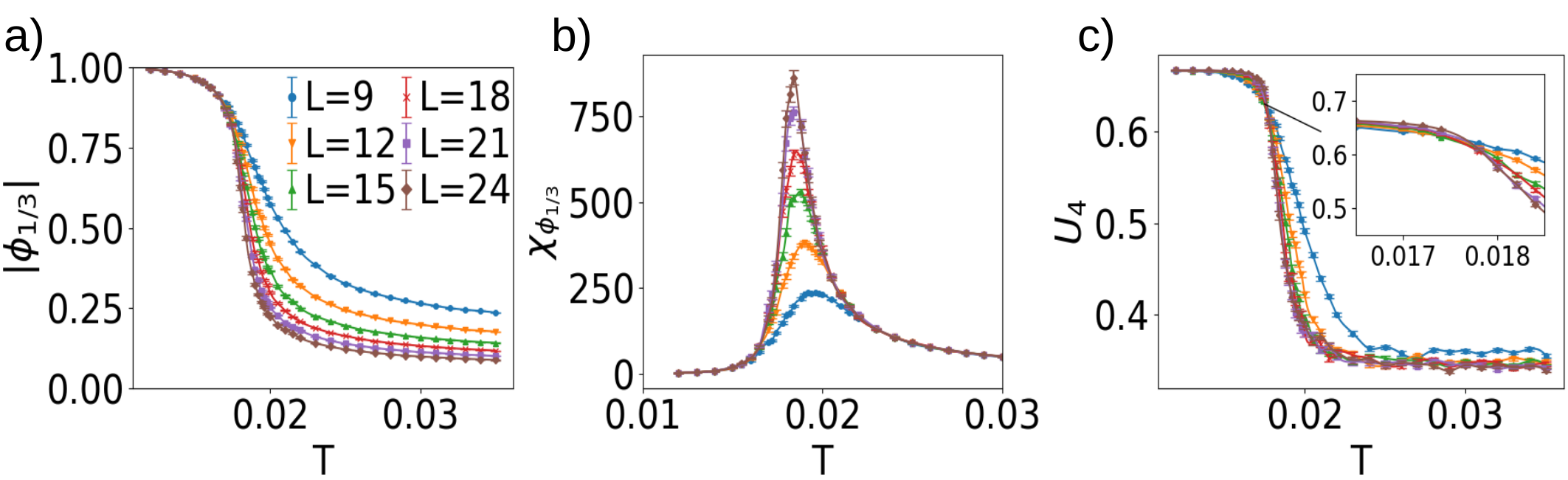} 
\caption{ Temperature driven transition for large $U$ ($U=30$). Order parameter's amplitude (a); Susceptibility (b); and Binder cummulant (c) plotted as a function of temperature for different systems sizes. The inset shows a zoom of the crossing point.
}
\label{transition}
\end{figure}

In order to characterize the temperature driven phase transition for large $U$ we have computed the order parameter susceptibility 
\begin{equation}
\chi_{\phi_{1/3}} = \beta V \left( \langle |\phi|^2 \rangle - \langle |\phi| \rangle^2 \right)
\end{equation} 
and the Binder cummulant
\begin{equation}
U_4 = 1 - \frac{\langle |\phi|^4 \rangle}{3\langle |\phi|^2 \rangle }.
\end{equation} 
These quantities are given in Fig.\ref{transition}.  
The crossing point of the Binder cumulant was used to determine the critical temperature.
Based on the finite size scaling of the susceptibility and the scaling of the derivatives with respect to $\beta=1/T$ of the Binder cumulant and the order parameter, we obtained the critical exponents quoted in the main text. The finite size scaling analysis  follows closely what is described in Ref. \cite{Fehske.10}. 

\section{Double occupancy} 

The number of doubly occupied sites is defined as 
\begin{equation}
D= \frac{1}{V}\sum_i \langle n_{f,i} c^\dagger_i c_i \rangle.
\end{equation} 
This quantity is explicitly computed as   
\begin{equation}
\langle n_{f,i} c^\dagger_i c_i \rangle = \sum_{\ \{ n_f \} }  n_{f,i} \left( \sum_\alpha |\langle i|\alpha \rangle|^2 n_F(\epsilon_\alpha)  \right)   e^{-\beta F(n_f)}
\end{equation} 
where $\ket\alpha $ is an eigenvector of the single-particle Hamiltonian $H(n_f)$ with energy $\epsilon_\alpha$ obtained for a given configuration of the classical charges. $n_F(\epsilon_\alpha)$ is the Fermi occupation number of the eigenstate $\ket\alpha$.

\begin{figure}[t!]
\centering
\includegraphics[width= 0.5 \textwidth]{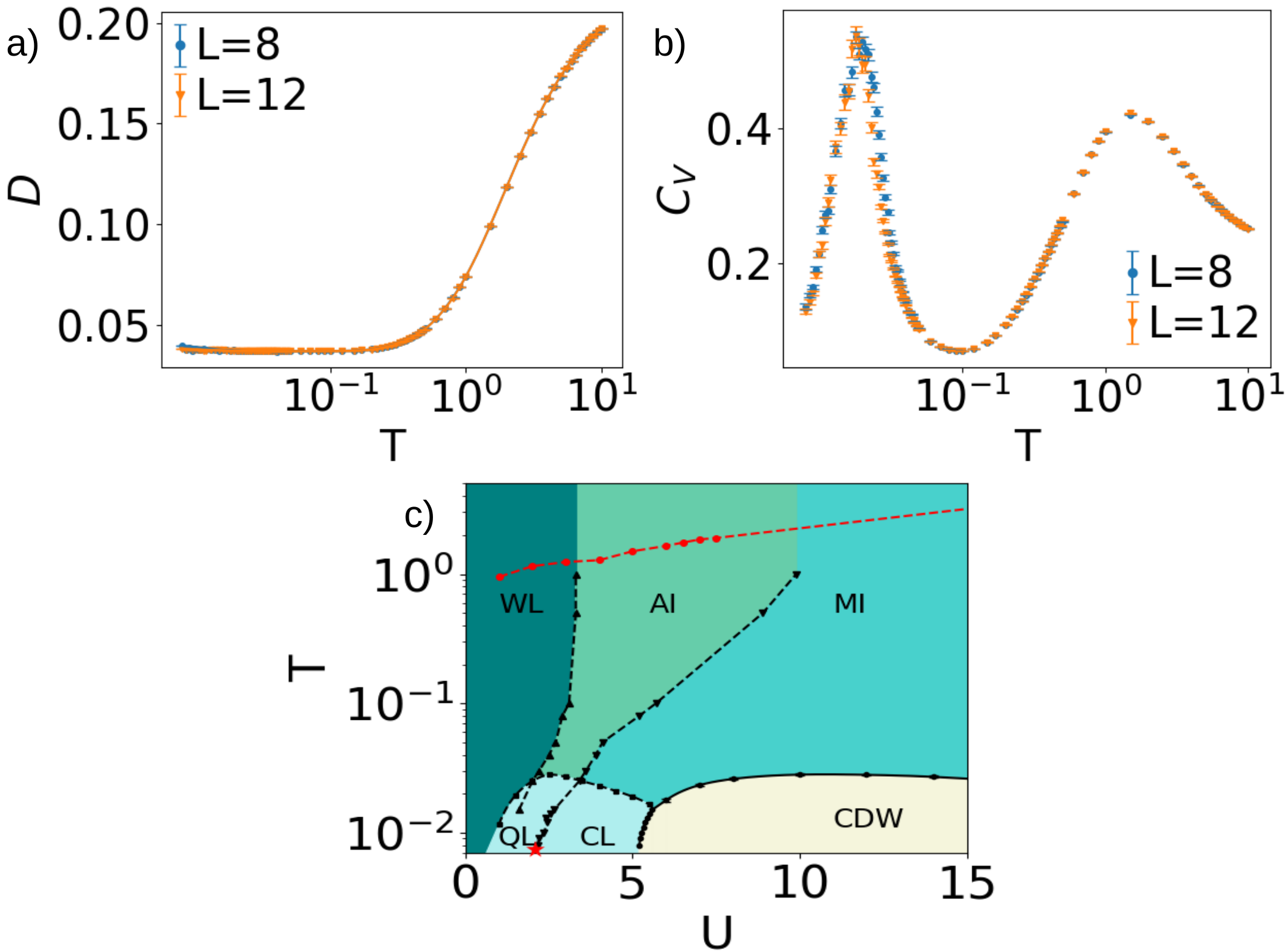} 
\caption{ Double occupancy (a) and specific heat (b) as a function of temperature for $U=4$. Phase diagram (c)  where the red line marks the position of the high temperature bump in the specific heat, which is associated with the suppression of double occupancy.
}
\label{double}
\end{figure}

Figs. \ref{double} (a) and (b) show that the high temperature peak of the specific heat coincides with the abrupt suppression of the double occupancy.
Fig. \ref{double} (c) depicts in a red line the maximum of the high temperature peak in the $C_v$. 
The low temperature phases studied in the main text arise for temperatures one order of magnitude lower than the sudden suppression of doubly occupied states and thus the phenomena reported in the main text cannot be attributed to it.

\section{Nature of the CL-CDW phase transition}

Our numerical results indicate that the transition between the CL and CDW phases, shown in the phase diagram of Fig.1-(a) (main text), is a first order transition.  

\begin{figure}[t!]
\centering
\includegraphics[width= 0.5 \textwidth]{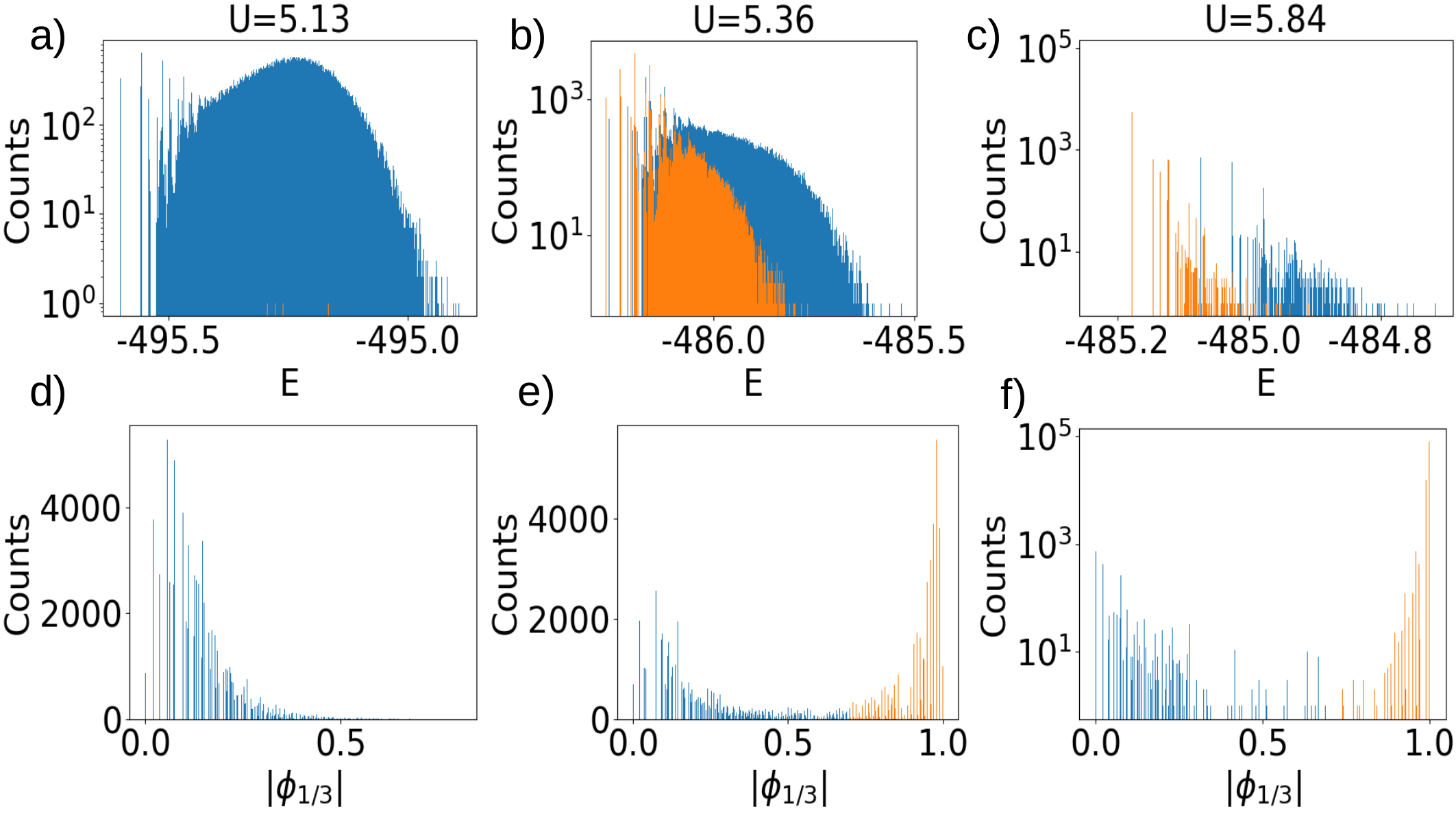} 
\caption{ Histograms of the energy (a,b,c) and of the order parameter (d,e,f) taken at different values of $U$ across the CL-CDW phase transition for $L=12$. Color coding: blue corresponds to all values of $|\phi_{1/3}|$ and orange to $|\phi_{1/3}|>0.7$.  
}
\label{discon}
\end{figure}

\begin{figure}[t!]
\centering
\includegraphics[width= 0.5 \textwidth]{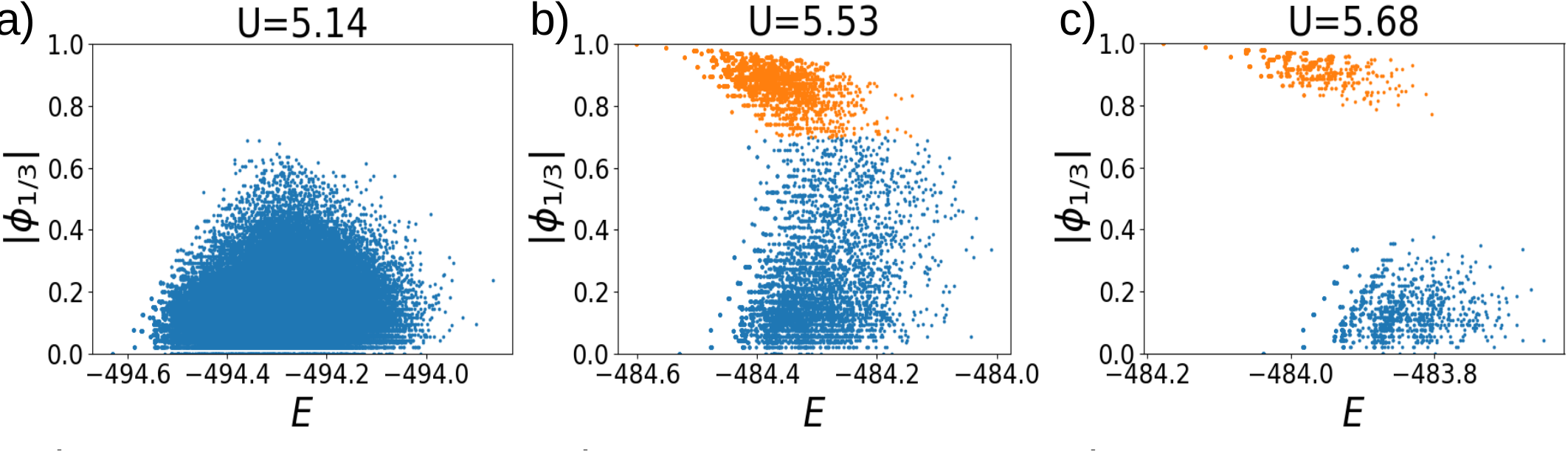} 
\caption{Amplitude of the order parameter, $|\phi_{1/3}|$ versus energy, $E$, for 100000 thermalized configurations for several values of $U$ taken at $T=0.011$ and $L=12$. The color coding is the same as in Fig.\ref{scatter}.
}
\label{scatter}
\end{figure}

Figs. \ref{discon} (a,b,c) and  \ref{discon} (d,e,f) show the histograms of the energy (a,b,c) and of the order parameter (d,e,f) for different values of $U$ across the transition. 
Figs. \ref{scatter} (a,b,c) show the distribution of configurations in the energy-order-parameter plane. 

A bipartite distribution of the order parameter appears in the vicinity of the transition. This can be seen in Figs. \ref{discon} (b) and (c) for the energy distribution, where a double peaked structure signals  the existence of two local minima that exchange stability at the transition, as well as in the distribution of the order parameter, Figs.\ref{discon} (e) and (f). 
The bipartite nature of the distribution can be read off more clearly from Fig. \ref{scatter} (c) where is color coding is used for configurations with different values of $|\phi_{1/3}|$, to highlight this feature.    
This hallmark of a discontinuous  transitions indicates that the CL-CDW transition is of first order.  

\begin{figure}[t!]
\centering
\includegraphics[width= 0.3 \textwidth]{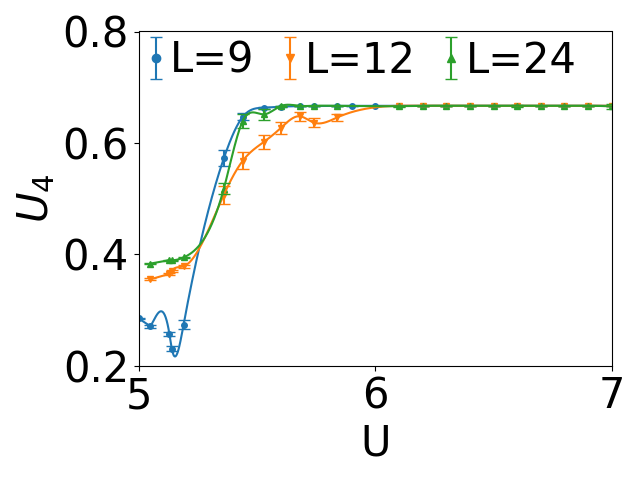} 
\caption{ Binder cumulant as a function of $U$ computed across the CL-CDW phase transition for $T=0.011$ and several system sizes. 
}
\label{binder}
\end{figure}

Fig. \ref{binder} depicts the Binder cumulant across the CL and CDW phase transition. For  system sizes that are commensurate with excitations corresponding to fillings $1/4$ and $1/3$ there is no crossing of the Binder cumulant. 
Although, $L=9$ crosses the $U_4(U)$ curves for $L=12$ and $L=24$, the lack of commensurability with these systems sizes makes a meaningful comparison difficult. 
It is worth noting that, in contrast with typical temperature-driven first order transitions, the Binder cumulant does not become negative in the vicinity of the transition. 
Therefore the negativity of the Binder cumulant, generally taken as being an indicator of a of first-order transition, cannot be used in this case. 
To which extend this finding points to a transition that is only weakly first order or even a continuous transition in the thermodynamic limit or simply a reflection of particular finite-size effects, is an open question.
%

\section{Phase separation} 

As mentioned in the main text, the liquid phase is characterized by a susceptibility $\chi(Q)$ which is enhanced for the same wave-vectors as those of the ordered phase that ensues for fillings $x_f=1/4$ and $x_c=3/4$. The order parameter of that phase is given by
\begin{equation}
\phi_{1/4} = \frac{4}{V} \sum_i \left[ e^{ 2\pi i} \delta_{A,i} + e^{ \frac{\pi}{2} i} \delta_{B,i} + e^{ \pi i} \delta_{C,i} + e^{ \frac{3 \pi}{2} i} \delta_{D,i}  \right] n_{f,i},
\label{order_4}
\end{equation}
where 
$A,B,C$ and $D$ refer to the 4 different sub-lattices and
$\delta_{A,i}$ is a function that is 1 when $i$ belongs to sub-lattice $A$ and 0 when it does not. Since an ordered phase of this type would only occupy one of the four sub-lattices, this order parameter is able to distinguish between those four different ground-states. 
As already stressed in the main part of the manuscript, this type of order is absent in the CL phase which follows from $\phi_{1/4}$ vanishing as $1/L$. This also rules out the occurrence of phase separation with  a non-vanishing volume fraction of a $\phi_{1/4}$ phase.
Interestingly, the effective classical model obtained by truncating the $t/U$ expansion to fourth order does exhibit phase separation with a $\phi_{1/4}$ phase.  

\begin{figure}[t!]
\centering
\includegraphics[width= 0.5 \textwidth]{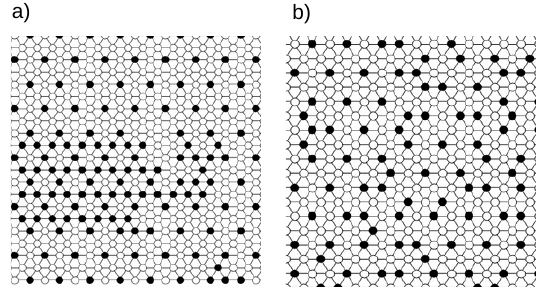} 
\caption{ 
(a) Configuration of an effective classical model with only a square plaquette term. 
(b) Typical configuration of the FKM within the  CL phase for $U=5$ and $T=0.0085$.
}
\label{separation}
\end{figure}

Fig. \ref{separation}(a) and (b) show two typical configurations of the classical charges for the phase separated case and for the CL phase. The dissimilarities between the two cases are self-evident and are indicative of the different nature of the two phases.

\section{Effective Classical Model} 

In this section we outline the derivation of the effective classical model, obtained from an expansion in small $t/U$, at finite temperature. Results for zero temperature were already given in ref. \cite{Gruber1997}.

For simplicity, the representation was   changed from occupation to spin variables $s_i = 2 n_{f,i} - 1$. 
Next, we expand the free energy in terms of $t/U$,
\begin{align}
F(s)& =F_0(s) + \frac{1}{\beta} \sum_{k=1}^\infty \frac{(-1)^k}{k}  \text{Tr}\left[(G_0T)^k\right] 
\end{align}
where
\begin{equation}
F_0(s) = -\mu_f N_f - \frac{1}{\beta} \text{Tr} \left[\log \left(- G_0^{-1}\right) \right], 
\end{equation}
$T$ is the adjacency matrix defined as $T_{i,j}=t$ if $i$ and $j$ are nearest neighbours on the lattice and 
\begin{equation}
G_0^{-1}(i\omega_n)= i\omega_n - \frac{U}{2} (S+1) + \mu_c.
\end{equation}
where $S$ is a diagonal matrix whose entries are the spin variables $S_{i,j}=\delta_{i,j} s_i$.

After summing over the Matsubara frequencies and going over rather tedious combinatorics, we obtain up to 4$^{\mbox{\tiny th}}$ order and for $T=0$
\begin{align}
E &=-\frac{1}{2}\left(\mu_f-\mu_c+\frac{6t^3}{U^2}\right)\sum_i s_i  + \left[ \frac{t^2}{2U}-\frac{6t^4}{U^3} \right] \sum_{|i-j|=1}s_is_j+ \nonumber \\
& +\frac{3t^3}{2U^2} \sum_\Delta s_\Delta +  \frac{5t^4}{2U^3}\sum_P s_P +\frac{3t^4}{2U^3} \sum_{|i-j|=\sqrt{3}} s_is_j +  \nonumber  \\ 
&+ \frac{t^4}{U^3} \sum_{|i-j|=2} s_is_j + O\left(\frac{t^5}{U^4} \right).
\label{truncated_classical}
\end{align}
These coefficients coincide with those given in Ref. \cite{Gruber1997}. 
The finite temperature expressions can  be obtained in a similar  way but will not be reproduced here as  they are lengthy and not particularly instructive. 
In Eq.~(\ref{truncated_classical}), the $\sum_{|i-j|=1}s_is_j$ term corresponds to an Ising-like interaction over nearest neighbors, $\sum_{|i-j|=\sqrt{3}}s_is_j$ over opposite ends of a rhombus and $\sum_{|i-j|=2}s_is_j$ over next-nearest neighbors. $\sum_\Delta s_\Delta$ corresponds to a sum over interactions on triangular plaquettes, where $s_\Delta$ is a product of the spins belonging to it and $\sum_P s_P$ represents the same for square plaquettes. A sketch of all the mentioned interactions is shown in the inset of Fig. Fig 4-(b) of the main text.

\section{Variational Classical Model} 

The variational model referred to in the main text was obtained assuming a free energy of the form of Eq.(\ref{truncated_classical}) but with arbitrary coefficients $J_{a=1,2..}$ 
\begin{align}
E &= h\sum_i s_i  + J_1 \sum_{|i-j|=1}s_is_j + J_2\sum_\Delta s_\Delta + \nonumber \\
& + J_3\sum_P s_P +J_4\sum_{|i-j|=\sqrt{3}}s_is_j + J_5\sum_{|i-j|=2}s_is_j.
\label{variational_classical}
\end{align}

\begin{figure}[H]
\centering
\includegraphics[width= 0.4 \textwidth]{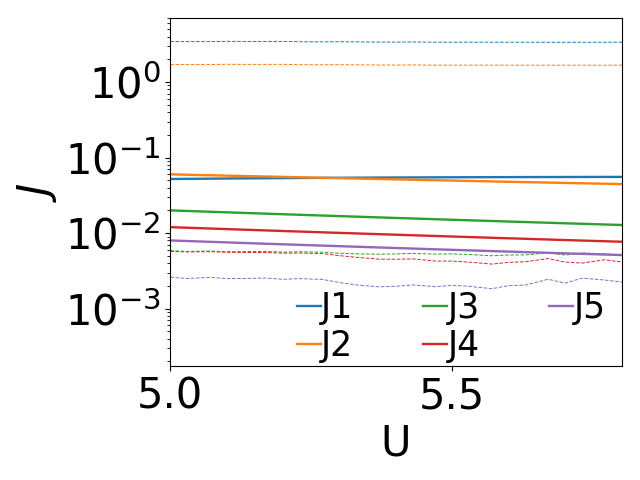} 
\caption{ 
Variational coupling constants as a function of $U$. The thin dashed lines are the values obtained with the linear regression while the bold lines are obtained from the expansion at $T=0$. 
}
\label{coup}
\end{figure}

The couplings were fixed considering $100000$ configurations of the classical charges generated with the FKM and fitting the free energy to that of Eq.(\ref{variational_classical}) with a linear regression on the $J_{a=1,2..}$  parameters. 
This procedure was done in Ref.\cite{Liu.17} to show that an effective Hamiltonian obtained via linear regression can be used to propose more efficient Monte-Carlo updates and thus speed up the dynamics of the original model.
The variational coupling constants obtained in this way are shown to vary smoothly with $U$ in Fig.\ref{coup}.

\end{document}